\documentclass[conference]{IEEEtran}
\IEEEoverridecommandlockouts
\usepackage{cite}
\usepackage{amsmath,amssymb,amsfonts}
\usepackage{algorithmic}
\usepackage{graphicx}
\usepackage{subcaption}
\usepackage{textcomp}
\usepackage{xcolor}
\usepackage{float}
\usepackage{adjustbox}
\usepackage{rotating}
\usepackage{url}

\def\BibTeX{{\rm B\kern-.05em{\sc i\kern-.025em b}\kern-.08em
    T\kern-.1667em\lower.7ex\hbox{E}\kern-.125emX}}

\newcommand{\Bpara}[1]{\vspace{1mm}\noindent{\bf #1.}}

\begin{document}

\title{Spatial Encoding of Flow Spaces \\ for Intelligent SDN Applications

% \title{Flow Space Encoding for SDN Applications
%\title{Spatially-Aware Flowset Encoding for Intelligent SDN Applications
%\title{Spatially-Aware Bloom Filter Based Flowset Encoding for Intelligent SDN Applications
\thanks{© 2025 IEEE. Personal use of this material is permitted. 
Permission from IEEE must be obtained for all other uses, 
in any current or future media, including reprinting/republishing 
this material for advertising or promotional purposes, 
creating new collective works, for resale or redistribution 
to servers or lists, or reuse of any copyrighted component 
of this work in other works.}

\thanks{This work is supported in part by U.S. NSF award 2346681.}
}

% \thanks{This work was supported in part by the U.S. National Science Foundation (NSF) under Grant XYZ-12345. \\
% © 2025 IEEE. Personal use of this material is permitted. Permission from IEEE must be obtained for all other uses, in any current or future media, including reprinting/republishing this material for advertising or promotional purposes, creating new collective works, for resale or redistribution to servers or lists, or reuse of any copyrighted component of this work in other works.}

% \author{
% \IEEEauthorblockN{Abdur Rouf}
% \IEEEauthorblockA{
% \textit{Computer Engineering} \\
% \textit{University of Central Florida} \\
% Orlando, USA \\
% abdur.rouf@ucf.edu}
% \and
% \IEEEauthorblockN{Murat Yuksel}
% \IEEEauthorblockA{
% \textit{Computer Engineering} \\
% \textit{University of Central Florida} \\
% Orlando, USA \\
% murat.yuksel@ucf.edu}
% }
\author{Abdur Rouf and Murat Yuksel\\
%Department of Computer Engineering, 
University of Central Florida, Orlando, FL, USA\\
\{abdur.rouf, murat.yuksel\}@ucf.edu}

\maketitle

\begin{abstract}
Efficient encoding of network flow spaces while preserving spatial locality is essential for intelligent Software-Defined Networking (SDN) applications, particularly those employing reinforcement learning (RL) methods in a reactive manner. In this work, we introduce a spatially aware Bloom Filter-based approach to encode IP flow pairs, leveraging their inherent geographical locality. Through controlled experiments using IoT traffic data, we demonstrate that Bloom Filters effectively preserve spatial relationships among flows. Our findings show that Bloom Filters degrade gracefully, maintaining predictable spatial correlations critical for RL state representation. We integrate this encoding into a DQN-based eviction strategy for reactive SDN forwarding. Experiments show that Bloom Filter-encoded, spatially aware flow representation enables up to 7\% and 8\% reduction in normalized miss rate over LRU and LFU, respectively, across 10 hours of traffic, demonstrating potential for low-latency applications. This experiment justifies the usefulness of preserving spatial correlation by encoding the flow space into a manageable size, opening a novel research direction for RL-based SDN applications.
\end{abstract}

% \begin{IEEEkeywords}
% Software-Defined Networking (SDN), Deep Reinforcement Learning, Bloom Filters, Flowset Encoding, Spatial Locality, DQN-Based Eviction
% \end{IEEEkeywords}

\section{Introduction}
% \begin{itemize}
% \item Brief background on SDN challenges related to flow management.
% \item Motivation: Importance of efficient flowset encoding in SDN.
% \item Reinforcement learning in SDN and why it matters.
% \item Why it is crucial for RL-based SDN applications to have a scalable/efficient representation of the flows in the environment.
% \item Research gap and our specific research question.
% \item Brief introduction of our method and contributions.
% \end{itemize}

Software-Defined Networking (SDN) has revolutionized the way modern data centers and networks are managed by decoupling the control plane from the data plane. This architectural shift allows for centralized network control, programmability, and flexibility, making SDN a foundational technology in cloud infrastructure, enterprise networks, and edge computing environments~\cite{sdn_survey}.
Despite its advantages, SDN faces critical challenges in flow management. A core component of SDN is the controller, which governs the behavior of multiple software or hardware switches using protocols such as OpenFlow~\cite{openflow}. These switches can observe millions of unique packet flows due to high network utilization, but they are typically constrained to install only a few thousand flow rules at a time. This limitation arises from the use of Ternary Content Addressable Memory (TCAM), a high-speed yet expensive memory optimized for rapid packet-field matching~\cite{tcam_limitations}. 
%TCAM is significantly more costly and power-hungry than conventional memory like DRAM or SRAM, and this economic constraint limits the number of flow entries that can be supported~\cite{tcam_limitations}. 
Furthermore, modern SDN applications often require rules with 16–20 match fields, each comprising multiple bits. The combinatorial explosion of these fields renders exhaustive rule generation computationally infeasible \cite{yu2010scalable}.

To support complex use cases such as Distributed Denial of Service (DDoS) mitigation, anomaly detection, and traffic classification (e.g., distinguishing mice and elephant flows), the SDN environment must capture rich and diverse flowset information~\cite{ddos_detection, mice_elephant}. Recent approaches have turned to machine learning (ML), deep learning (DL), and reinforcement learning (RL) to address these challenges~\cite{rl_sdn}. These methods rely on high-quality input representations of flow behavior to make effective decisions. A particularly important relationship within a flowset is the \textit{spatial locality} that flows often exhibit patterns based on source-destination geographical or topological proximity. Preserving such relationships in the input state is essential for RL agents to learn and adapt effectively in dynamic network conditions.
However, maintaining this spatial information while providing a scalable and compact state representation for RL-based SDN applications remains a significant challenge. 
RL agents such as Deep Q-Network (DQN), Double DQNs, and Proximal Policy Optimization (PPO) are sensitive to state dimensionality and quality. An uncompressed flowset with full spatial detail may not fit into the limited memory of the controller or meet the time constraints required for real-time inference. This necessitates an efficient and loss-tolerant encoding technique.

In this study, we propose a novel Bloom Filter-based encoding strategy that preserves spatial locality while significantly reducing the size of flowset representations. Bloom Filters, known for their space efficiency and probabilistic nature, offer a graceful degradation property that aligns well with the robustness needs of RL systems. To the best of our knowledge, this is the first work that utilizes Bloom Filters not just for compression, but also as a mechanism to preserve spatial relationships in flow spaces for intelligent SDN applications.

Through controlled experiments on real-world IoT traffic traces, we demonstrate that our encoding method maintains useful spatial correlations among flows and supports effective learning in a DQN-based caching mechanism. Our results show that Bloom Filter-based encoding enables the RL agent to reduce the normalized miss rate by approximately 7\% and 8\% compared to Least Recently Used (LRU) and Least Frequently Used (LFU) eviction strategies, respectively, over 10 hours of observed traffic. This confirms the efficacy of our method in enabling low-latency, high-performance decision-making in resource-constrained SDN switches.
Overall, this work opens a new direction in flow representation for RL-based SDN systems by introducing spatially-aware encoding that balances fidelity and scalability. We believe this approach can generalize to a broader class of learning agents and network scenarios.

\section{Motivation: Flow Set Encoding in SDN}

% \begin{itemize}
% \item Definition and significance of flowset encoding in network management: Security applications such as identifying flows that are part of a DDoS attack; QoS applications such as separation of elephant and mice flows.
% \item Why SDN applications face a huge challenge in expressing the entire set of flows: 16 fields generate a very large flow space?
% \item Why RL/ML agents in SDN need a more granular and scalable way of representing the flowset in the environment? -- the need for flowset encoding
% \end{itemize}

In SDN environments, the term \textit{flowset} refers to the collection of packet flows or flow rules observed within a network over a period of time. Encoding this flowset efficiently is critical for a range of network management tasks, especially those that require real-time decision-making. For example, security applications such as detecting or mitigating DDoS attacks often rely on identifying patterns across multiple flows originating from diverse sources targeting a single victim~\cite{ddos_detection}. Similarly, Quality-of-Service (QoS) policies require distinguishing between high-bandwidth ``elephant'' flows and latency-sensitive ``mice'' flows, necessitating a complete yet manageable view of the active flows~\cite{mice_elephant}. Without a compact representation of the flow space, applications must operate on raw or partial flow data, limiting their effectiveness and scalability in high-throughput networks.

One of the fundamental limitations in SDN stems from the expressiveness and size of the flowset. Modern SDN controllers define flows using multiple match fields—such as source/destination IP addresses, ports, protocol type, VLAN ID, TCP flags, and more—many of which span 10s of bits each. The OpenFlow standard, for instance, supports up to 16 or more match fields~\cite{openflow}. The Cartesian product of all possible values for these fields creates a combinatorially explosive flow space, far beyond what can be stored, computed, or even reasoned about exhaustively. This challenge is further exacerbated by the dynamic nature of network traffic, where flows may appear, evolve, and disappear rapidly. Attempting to enumerate or monitor the entire set of possible flows in real time is, therefore, computationally infeasible for both the controller and the underlying monitoring infrastructure.

To manage this complexity, many modern SDN applications leverage RL and ML agents to make intelligent decisions about routing, caching, and resource allocation. However, these agents rely heavily on the input representation of the environment—i.e., the flow space —to learn effective policies. A high-fidelity yet compact state is essential for enabling these agents to generalize and make accurate decisions. Feeding the entire flowset as raw input into an RL agent is impractical due to memory constraints and training inefficiencies. Moreover, important flow relationships—especially those with spatial relevance, such as locality patterns in IP source-destination pairs—can be lost if the flowset is sampled or encoded naively. This creates an urgent need for encoding techniques that preserve essential characteristics, such as spatial location, while reducing the dimensionality of the state space~\cite{rl_sdn, rl_sdn_2}. An effective flowset encoding can thus serve as a bridge between the raw network data and the abstract, learnable representations required by RL agents, enabling scalable and adaptive SDN control.

\section{Related Work}

% \begin{itemize}
% \item Overview of Bloom Filters and their typical applications.
% \item Existing SDN flow management/caching methods (e.g., LRU-based, LFU and hybrid).
% \item Existing reinforcement learning approaches in SDN for flow caching.
% \item Limitations or gaps identified in existing literature.
% \end{itemize}

%\subsection{Bloom Filters in Networked Systems}
\Bpara{Bloom Filters in Networked Systems}
Bloom Filters are probabilistic data structures used for efficient membership testing with minimal memory overhead~\cite{bloom1970space}. A Bloom Filter represents a set as a fixed-size bit array and multiple independent hash functions. When inserting an item, such as an IP flow pair (e.g., \texttt{192.168.1.1:1234 $\rightarrow$ 10.0.0.1:80}), each of the $k$ hash functions maps the item to a position in the bit array and sets the corresponding bits to 1. To check if a given flow is in the set, the same hash functions are applied to the query item, and if all corresponding bits are set to 1, the item is considered ``possibly present." Otherwise, it is definitely not present. Due to hash collisions, Bloom Filters may return false positives—reporting presence of items that were not inserted.

In the context of networking, Bloom Filters have been widely used in areas such as packet routing~\cite{bloom_routing}, intrusion detection~\cite{bloom_ids}, and network measurement~\cite{bloom_measurement}. Their ability to compress large sets with tunable error rates makes them particularly useful for memory-constrained environments such as edge routers and IoT devices. In our work, we utilize Bloom Filters to generate bit arrays from sets of flow rules. These bit arrays are then used as spatially aware state representations for RL agents in SDN environments.
%, which we elaborate on in later sections.

\Bpara{Flow Management and Caching in SDN}
Flow management is a central task in SDN, as the forwarding tables in data plane switches are limited by the size and cost of TCAM. To maximize utility, many systems adopt flow caching strategies that dynamically insert and evict flows from these tables based on observed traffic patterns. Classic approaches such as LRU and LFU are widely adopted for this purpose~\cite{lru_sdn, lfu_sdn}. LRU evicts the flow that has not been accessed for the longest time, while LFU removes the least accessed flow. Hybrid approaches (e.g., LRU-LFU combinations) aim to balance recency and frequency, achieving higher hit rates in diverse traffic conditions ~\cite{lfu_sdn}. 
These caching strategies are computationally lightweight and have shown good empirical performance in many scenarios. By efficiently utilizing limited memory, they can maintain high cache hit rates. However, their heuristic nature means they may fail to adapt optimally under rapidly changing traffic or context-aware scenarios.

To improve upon static heuristics, researchers have proposed the use of RL to guide flow caching decisions. These methods leverage traffic patterns and historical behaviors to learn optimal cache eviction policies. DQN-based~\cite{rl_dq_caching, rl_dq_caching_2}, Actor-Critic~\cite{rl_actorcritic}, and PPO-based~\cite{rl_ppo_sdn} techniques have been used to dynamically adjust cache contents in real time. These RL-based methods have demonstrated significant improvements in cache hit rates over traditional techniques, especially under nonstationary or adversarial traffic conditions.

\Bpara{Flow Space Representation for RL}
Despite their potential, most RL-based SDN systems face a critical bottleneck: the state representation. As flow counts scale into the millions, feeding raw or flattened flow features into RL agents becomes computationally infeasible. Some studies attempt to cluster or downsample flows~\cite{flow_clustering_rl}, but often at the cost of losing important spatial or temporal relationships. This limitation results in suboptimal learning and decision-making, particularly in applications that require fine-grained flow-level control.

While RL-based flow management shows promise, most existing works do not adequately address the scalability of state representation. The lack of compact yet expressive encodings for large flow spaces impedes learning performance and system responsiveness. In particular, the spatial relationships among flows—such as those originating from similar subnets or destined to similar services—are often ignored or poorly modeled. This gap limits the potential of RL-based SDN systems in environments where flow behavior is highly dynamic and context-sensitive.
Our work addresses this limitation by proposing a Bloom Filter-based encoding that preserves spatial locality while significantly reducing the state size. This approach enables efficient, scalable flow representation suitable for RL-driven SDN controllers, paving the way for more adaptive and intelligent network management.

% \section{Design Approach and Problem Statement}

% \subsection{Parameters}
% \begin{table}[htbp]
% \centering
% \begin{tabular}{|l|l|l|}
% \hline
% \textbf{Parameter} & \textbf{Range} & \textbf{Purpose} \\ \hline
% Window sizes & 30-290 (step 20) & Test different data locality \\ \hline
% False positive rates & 5-30\% (step 5\%) & Bloom Filter accuracy \\ \hline
% Flip percentages & 5-30\% (step 5\%) & Robustness \\ \hline
% Seeds & First 1000 primes & Reproducible randomness \\ \hline
% \end{tabular}
% \label{tab:experiment-parameters}
% \end{table}

% \subsection{Outputs}
% \begin{table}[htbp]
% \centering
% \begin{tabular}{|l|l|}
% \hline
% \textbf{Objectives} & \textbf{Short Description} \\ \hline
% False Positive Count & Normalized false positive count w.r.t the flow set size \\ \hline
% Mean Distance & Distance based on concatenated value-based sorting \\ \hline
% True Positive Count & \begin{tabular}[c]{@{}l@{}}Normalized true positive count w.r.t selected flows \\ initially inserted to generate the Bloom Filter bit array\end{tabular} \\ \hline
% \end{tabular}
% \label{tab:experiment-outputs}
% \end{table}

% \begin{itemize}
% \item Clear statement of the research problem.
% \item Design principles behind spatially-aware encoding.
% \item Why spatial locality preservation matters for RL agents.
% \item How our approach uniquely addresses existing limitations.
% \item Expected advantages over traditional encoding methods.
% \end{itemize}

\section{Problem Statement and Design Approach}

%\subsection{Problem Statement}

In RL-based SDN systems, providing an expressive yet compact state representation is essential for effective decision-making. However, capturing millions of concurrent flows while preserving meaningful relationships among them, such as spatial locality, poses a significant scalability challenge. Existing approaches either ignore inter-flow relationships or rely on high-dimensional raw input representations that overwhelm RL agents. This motivates our central research problem: \textit{Can we design a space-efficient encoding that preserves spatial locality within flow spaces to serve as a practical state representation for RL agents in SDN environments?}

%\subsection{Design Principles and Motivation}

We address two key challenges: (1) reducing the dimensionality of the flowset state representation, and (2) preserving spatial locality among flows. To this end, we evaluate the use of Bloom Filters as a candidate encoding strategy due to their lightweight memory footprint, probabilistic membership capabilities, and graceful degradation under perturbation. Our hypothesis is that spatially aware encoding using Bloom Filters can retain local clustering behavior among flow rules while offering a fixed-size representation suitable for RL agents.

\subsection{RL Integration with SDN: State, Reward, and Action Loop}

\begin{figure}[htbp]
    \centering
    \includegraphics[width=0.49\textwidth]{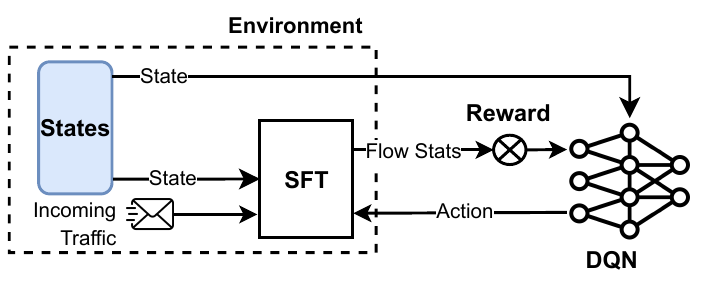}
    \vspace{-6mm}
    \caption{DQN-based SDN setup showing state, action, and reward flow: The Bloom Filter-based encoded flowset acts as the RL agent's input state.}
    \label{fig:dqn_sdn_overview}
    \vspace{-5mm}
\end{figure}

In our setup, the DQN agent interacts with the SDN environment in a closed feedback loop, as illustrated in \textbf{Figure~\ref{fig:dqn_sdn_overview}}. The environment consists of the incoming network traffic and the Switch Forwarding Table (SFT), which maintains active flow rules. The Bloom Filter-based flowset encoding is used to generate the state representation from active flows observed at the switch. This encoded bit array is then passed to the DQN agent as input.
Based on this state, the DQN agent decides an action, e.g., selecting which flow(s) to evict or prioritize for caching or which flow rules to install. Once the action is applied, the environment updates the SFT and collects statistics such as hit/miss rate, eviction success, or latency performance. These metrics are then processed into a scalar reward signal, which is passed back to the DQN agent for training. The continuous loop of state-action-reward enables the DQN to learn optimal policies over time.

The effectiveness of this pipeline depends on the quality and scalability of the state representation. A naive or high-dimensional representation would increase the learning complexity and impair generalization. Our Bloom Filter encoding addresses this by offering a fixed-length bit array that maintains spatial relationships between flows while allowing for compact representation. As a result, the RL agent receives a meaningful and scalable state, improving decision-making efficiency in dynamic traffic scenarios.

\subsection{Experimental Methodology}

Our evaluation is based on packet traces of device communications collected by the IoT Traffic Analytics Research Group at UNSW Sydney~\cite{unsw_iot_dataset}. These traces include over 500K packets, from which we extract over 1,000 unique IPv4 source-destination pairs. Each IP address is converted to a 32-bit integer and concatenated to form a 64-bit flow identifier. The flowset is then sorted by these values to expose spatial locality—flows with similar subnets are close after sorting.

\begin{figure}[htbp]
    \centering
    \vspace{-3mm}
    \includegraphics[width=0.50\textwidth]{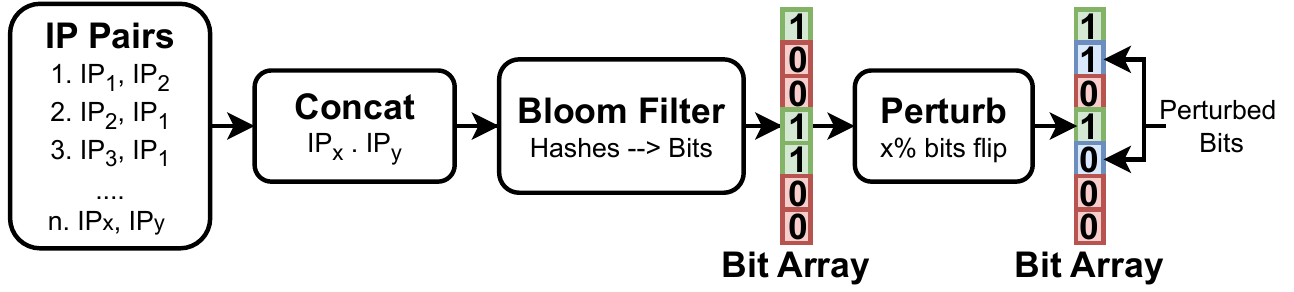}
    \vspace{-5mm}
    \caption{Bloom Filter-based encoding pipeline for flow-aware SDN: IP pairs are hashed into a bit array, perturbed, and used as input to a DQN agent.}
    \label{fig:bit_perturbation_pipeline}
    \vspace{-4mm}
\end{figure}

We simulate locality by choosing a random \textit{window start index} in the sorted list and varying the window size from 30 to 290 in steps of 20. Within each window, we randomly select half the flows (e.g., 15 from a window size of 30) to be inserted into a Bloom Filter. Each IP pair is transformed into a string format like \texttt{srcIP.dstIP} before insertion. The Bloom Filter is initialized using the total number of inserted items, $n$, and a target false positive rate, $P$, which determines the length of the bit array, $m$, using the formula:
$m = -\frac{n \ln P}{(\ln 2)^2}.$
After initialization, a percentage of bits in the bit array is flipped to simulate noise or perturbation, mimicking scenarios where the bit array may be affected by updates or approximations from an RL agent. This flip percentage ranges from 5\% to 30\% with 5\% increments.
\textbf{Figure~\ref{fig:bit_perturbation_pipeline}} illustrates the end-to-end encoding process that forms the state used in the DQN decision loop.

To assess the locality-preserving properties of the Bloom Filter, we analyze the distances between flows identified as positive (both true and false positives) and the center of the original window. For each identified positive flow, we record its index in the sorted global list and compute its absolute distance to the center index of the original window. We calculate and report the \textit{mean distance} for two categories:
\begin{itemize}
    \item \textbf{True Positives (TPs)}: Flows that were inserted into the Bloom Filter and still recognized after perturbation.
    \item \textbf{False Positives (FPs)}: Flows not inserted but marked present due to Bloom Filter's probabilistic nature.
\end{itemize}
This metric reflects the extent to which the encoded bit array continues to retain spatial locality information.

\subsection{Empirical Observations}

\begin{figure*}[htbp]
    \centering

    \begin{subfigure}[t]{0.24\textwidth}
        \centering
        \includegraphics[width=\textwidth, height=4cm, keepaspectratio]{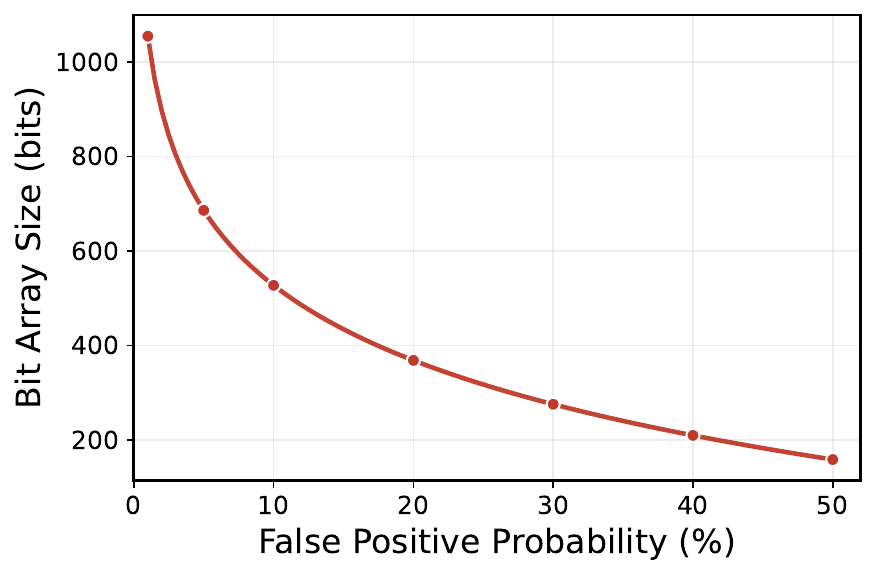}
        \caption{}
        \label{fig:bitarray_fp}
    \end{subfigure}%
    \hfill
    \begin{subfigure}[t]{0.24\textwidth}
        \centering
        \includegraphics[width=\textwidth, height=4cm, keepaspectratio]{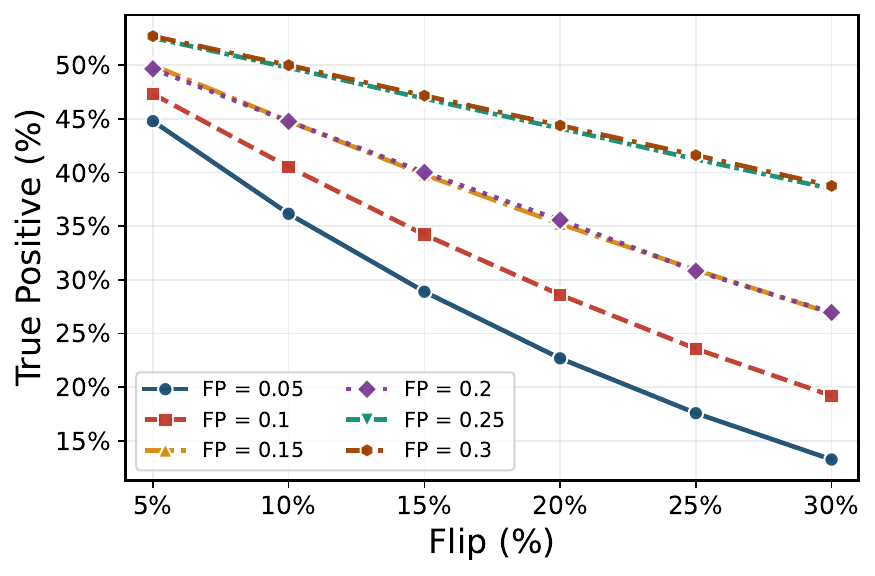}
        \caption{}
        \label{fig:tp_analysis}
    \end{subfigure}%
    \hfill
    \begin{subfigure}[t]{0.24\textwidth}
        \centering
        \includegraphics[width=\textwidth, height=4cm, keepaspectratio]{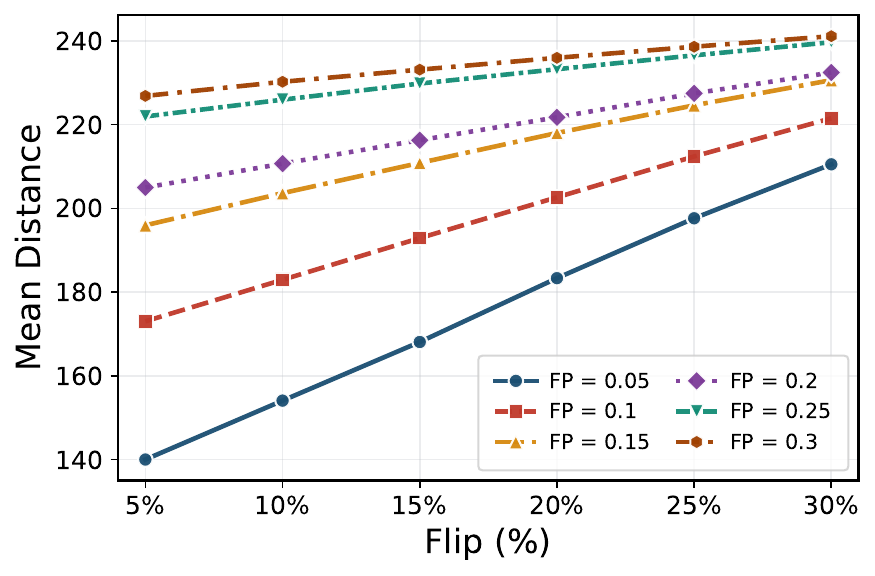}
        \caption{}
        \label{fig:distance_analysis}
    \end{subfigure}%
    \hfill
    \begin{subfigure}[t]{0.24\textwidth}
        \centering
        \includegraphics[width=\textwidth, height=4cm, keepaspectratio]{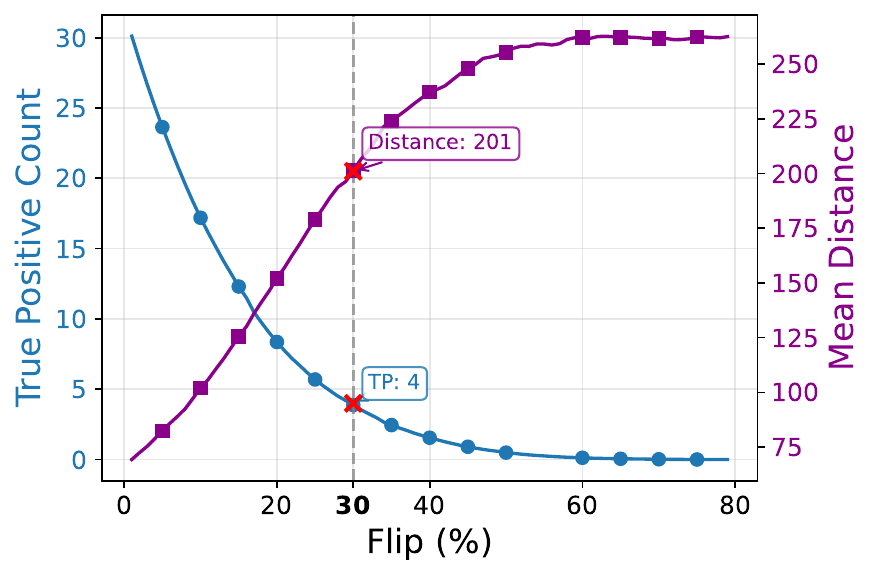}
        \caption{}
        \label{fig:bf_design_choice}
    \end{subfigure}
    \vspace{-2mm}
    \caption{Comprehensive Bloom Filter Analysis: (a) Bit array size vs. false positivity, (b) True positive rate across flip percentages, (c) Sensitivity to spatial locality (measured with a window of 110 and 55 inserted items), and (d) The effect of bit flip on true positives and average distance using a Bloom Filter with a window of 64 (32 inserted items) and a 1\% false positive rate.}
    \label{fig:full_bf_analysis}
    \vspace{-6mm}
\end{figure*}

Bloom Filters' bit array size is very sensitive to their False Positivity (FP) rate. \textbf{Figure~\ref{fig:bitarray_fp}} illustrates that the Bloom Filter bit array size decreases exponentially as the FP rate increases. This trend confirms the theoretical expectations where a higher FP rate reduces the number of bits required per inserted item.
For instance, lowering the FP rate from 30\% to 5\% increases the bit array size from under 200 bits to over 1,000 bits for a 55-item insertion. Based on this observation, we select a sweet spot in the FP range that balances two key objectives: maintaining a compact state representation and preserving useful visibility for RL. Using raw flow information as RL state would result in an exponentially large state space, making it computationally impractical. In contrast, encoding flowsets using Bloom Filters allows the RL agent to operate over a compact, fixed-size bit array that still retains valuable spatial structure.

% \textbf{Figure~\ref{fig:tp_analysis}} depicts how the true positive rate degrades with increased bit flipping. Filters with lower false positive rates (thus larger bit arrays) suffer steeper performance degradation, as a fixed flip percentage changes more absolute bits. This validates the trade-off between robustness and compression: smaller Bloom Filters are more resistant to perturbations at fixed flip ratios.

\textbf{Figure~\ref{fig:tp_analysis}} illustrates how the TP rate declines as the percentage of flipped bits increases. This behavior is expected, as bit flipping disrupts the original encoding, making it harder for the Bloom Filter to recognize flows it had previously inserted. However, at higher FP rates (i.e., smaller bit arrays), this degradation becomes less pronounced. Since the bit array is smaller, the same flip percentage alters fewer actual bits, which makes the filter more resistant to perturbation. This trade-off underscores how compact encodings offer better robustness at the cost of some precision, while larger bit arrays provide more accuracy but are more sensitive to bit-level perturbations.

% \textbf{Figure~\ref{fig:distance_analysis}} confirms that the false positive results, even after bit flipping, tend to be spatially close to the inserted window center. Lower false positive rates exhibit lower average distances, affirming that the Bloom Filter structure inherently retains some locality even under perturbation. This supports the hypothesis that spatial locality is preserved in the Bloom Filter encoding.

% \begin{figure*}[htbp]
%     \centering
%     \begin{subfigure}[t]{0.33\textwidth}
%         \centering
%         \includegraphics[width=\textwidth, height=4cm, keepaspectratio]{plots/bloom_filter_linear.pdf}
%         \caption{}
%         \label{fig:bitarray_fp}
%     \end{subfigure}%
%     \hfill%
%     \begin{subfigure}[t]{0.33\textwidth}
%         \centering
%         \includegraphics[width=\textwidth, height=4cm, keepaspectratio]{plots/true_positive_plot.pdf}
%         \caption{}
%         \label{fig:tp_analysis}
%     \end{subfigure}%
%     \hfill%
%     \begin{subfigure}[t]{0.33\textwidth}
%         \centering
%         \includegraphics[width=\textwidth, height=4cm, keepaspectratio]{plots/distance_plot.pdf}
%         \caption{}
%         \label{fig:distance_analysis}
%     \end{subfigure}
%     \caption{Comprehensive analysis across different false positivity: (a) Bit Array size vs. false positivity, (b) TP rate, and (c) Spatial locality sensitivity.}
%     \label{fig:complete_analysis}
% \end{figure*}

To further explore the behavior of Bloom Filters under perturbation, we extend our experiment by identifying which flows remain detectable after performing bit flips on the Bloom Filter bit array. For each perturbation level, we calculate the distance between the detected flows and the center of the original flow window (based on their sorted IP-pair integer values) and took the mean of these distances.
This deeper analysis led to an important finding: as the flip percentage increases, the Bloom Filter starts recognizing flows that are farther away from the original window center. In other words, \textit{more extensive perturbation causes the Bloom Filter to capture flows that are spatially more distant in the sorted flowset}. This confirms that Bloom Filters are sensitive to spatial locality — with higher bit alterations, the structure generalizes more broadly and includes distant flows, while with fewer bit changes, the filter remains more selective and local.
\textbf{Figure~\ref{fig:distance_analysis}} illustrates this phenomenon by showing that the average distance of FPs from the window center increases with flip percentage. Additionally, Bloom Filters configured with lower FP rates (and thus larger bit arrays) exhibit more sensitive spatial clustering around the window center, reinforcing the notion that locality is inherently preserved in the encoding process, even under noise.

We leverage this spatial sensitivity in our DQN-based eviction design. By encoding the flowset into a Bloom Filter and feeding the bit array into the agent, we provide a compressed yet spatially aware representation of the network state. Our subsequent experiments support the hypothesis that such spatially-informed states enable the DQN agent to make more precise and context-aware decisions, ultimately improving eviction accuracy and hit rate stability over time.

\subsection{Implications for RL-Based SDN Applications}

Our findings demonstrate that Bloom Filter-based flowset encoding not only compresses the flow state but also maintains spatial awareness, which is vital for RL agents learning flow behaviors. Notably, the degradation of spatial locality under bit perturbations is both gradual and consistent across the evaluated range of flip percentages and FP rates. This predictable degradation pattern is crucial because it allows the system to anticipate and adapt to noise or updates, which are common in dynamic RL environments. Leveraging this insight, we design a flow caching mechanism—discussed in later sections—that exploits this stability to make informed eviction decisions under RL-based policies. Compared to traditional representations (e.g., raw flow features or statistical summaries), our approach enables low-complexity state construction, preserves interflow spatial relationships, and enhances robustness, all while reducing the input dimensionality for learning agents.

\section{SDN Application: DQN-Based Flow Management}
\label{sec:DQN+LRU}

%In this section, 
We present our DQN-based flow eviction architecture that integrates Bloom Filter-based flowset encoding with RL for intelligent flow management in a reactive SDN environment.
\textbf{Figure~\ref{fig:dqn_flow_management}} provides a comprehensive illustration of our proposed %DQN-based flow eviction 
framework, explicitly designed to integrate reactive SDN mechanisms with intelligent eviction strategies. At the highest level, the architecture operates within a traditional reactive environment: when a packet arrives at a switch and no corresponding flow rule exists in SFT, the switch generates a \texttt{packet-in} message to the controller. In response, the controller processes the request and returns a \texttt{FlowMod} message containing the necessary flow rules. The latency associated with this operation is termed the \textit{Reactive Time Interval} (RTI).
In this context, the switch treats packets arriving during the RTI—before the flow rule installation—as misses. Conversely, packets matching existing flow rules in the SFT are counted as hits. Each installed flow rule includes an idle timer, which resets upon observing corresponding traffic. If no traffic matches the rule before the timer expires, the rule is removed automatically.

In scenarios where the SFT reaches full capacity,
%without idle rules, 
the LRU or LFU eviction policy is applied.
%traditionally selects a rule for removal. 
To enhance eviction efficiency and reduce packet misses, we introduce a DQN-based eviction mechanism, activated regularly at every \textit{Eviction Time Interval} (ETI). At each ETI, the system first evaluates the occupancy of the SFT. If the SFT is at maximum capacity, the following systematic process is executed:
\begin{figure*}[htbp]
    \centering
    \includegraphics[width=0.95\textwidth]{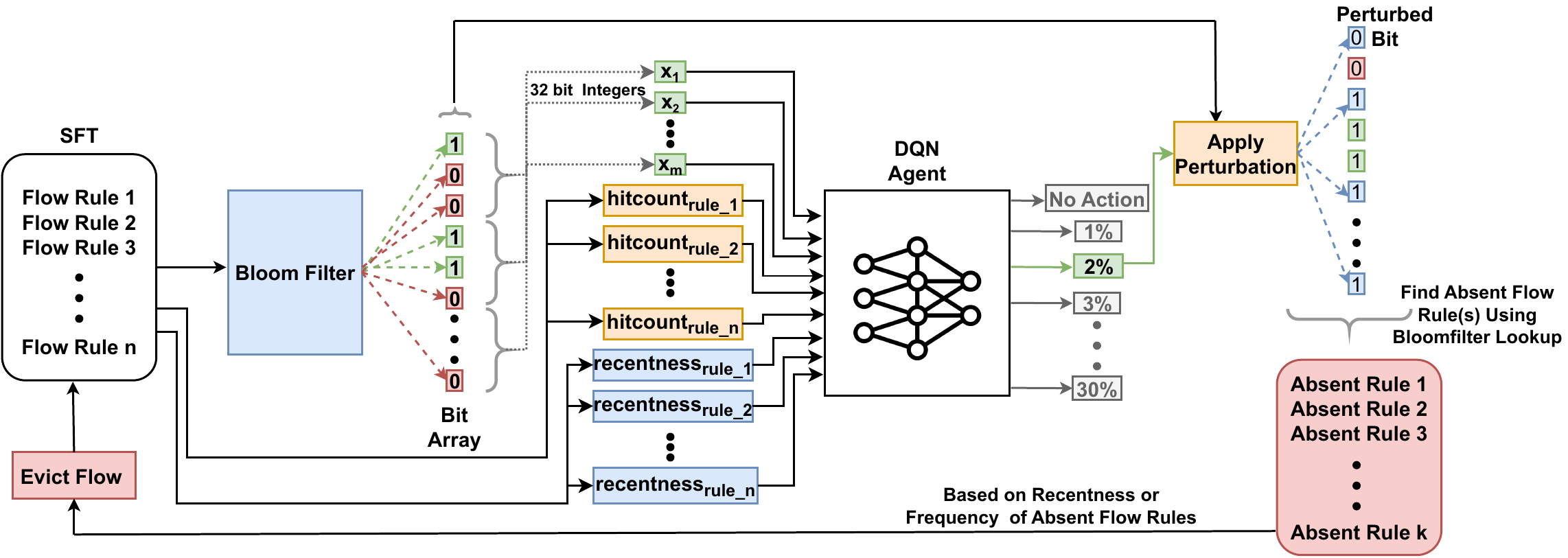}
    \vspace{-2mm}
    \caption{Architectural Overview: Integrating Bloom Filter-based encoding with DQN enables spatially-aware flow eviction by preserving and intelligently perturbing flow locality.}
    \label{fig:dqn_flow_management}
    \vspace{-6mm}
\end{figure*}

\begin{enumerate}
\item \textbf{Flow Rules Encoding:} The active flow rules, represented by concatenating their source and destination IP addresses, are encoded into a Bloom Filter. This step transforms the flowset into a compact bit array, effectively capturing spatial locality among flows.

\item \textbf{Bit Array Processing:} The resulting bit array is padded with leading zeros to ensure its length becomes divisible by 32, making it  separable neatly into 32-bit integers.

\item \textbf{Input State Formation:} These 32-bit integers, which encapsulate the spatial locality, along with flow metadata (i.e., hit counts and recentness of each flow rule) form the comprehensive input state for the DQN agent.

\item \textbf{Q-Value Calculation:} The DQN agent calculates Q-values from the provided state using a reward function: It awards a reward of +1 if the total hit counts increase compared to the previous ETI, assigns 0 if unchanged, and penalizes with -1 if decrease.

% \item \textbf{Action Selection and Perturbation:} Our DQN framework can choose from 31 distinct actions, encompassing no perturbation (0\%) and perturbations ranging from 1\% to 30\%. This range was specifically selected based on our experimental validation, which demonstrated that Bloom Filters reliably retain spatial locality information up to 30\% perturbation. Upon selecting a perturbation percentage, the corresponding fraction of bits in the Bloom Filter's bit array are randomly flipped, generating a perturbed bit array.
%\item \textbf{Action Selection and Perturbation:}
% \begin{figure}[h]
%     \centering
%     \includegraphics[width=0.30\textwidth]{plots/bf_design_choice.pdf}
%     \caption{Effect of bit flip percentage on TP count and average distance from window center. The Bloom Filter is configured with a window size of 64 (32 inserted items) and an FP rate of 1\%.}
%     \label{fig:bf_design_choice}
% \end{figure}

\item \textbf{Absent Flow Identification:} Using the perturbed Bloom Filter, the system performs look-ups against the current set of flow rules in the SFT. Flow rules unrecognized in this lookup are labeled as absent flows.

\item \textbf{Flow Eviction Decision:} Finally, among these absent flows, the rule with the lowest recentness, indicating the oldest traffic, or the lowest frequency, representing flows with fewer packet matches, is prioritized for eviction, thereby effectively freeing up SFT capacity for a new incoming flow.
\end{enumerate}

Our DQN agent selects from 31 discrete actions, ranging from 0\% (no perturbation) to 30\% bit perturbation in 1\% increments. This action space is  determined through empirical analysis of Bloom Filter behavior under varying levels of perturbation. To tune this range, we conduct an experiment where 32 items were inserted into a Bloom Filter, configured with a window size of 64 and a 1\% FP rate—ensuring approximately 10–12 bits per item, which is still compact compared to a typical string-based flow identifier. We then progressively flip 1\% to 80\% of the Bloom Filter's bits and observe the number of TPs retained and the mean distance of these matches from the center of the original window in the sorted flow space. In \textbf{Figure~\ref{fig:bf_design_choice}}, we observe that, at 30\% flip, about 12\% of the original items remain detectable and the mean distance approaches 200 when the total number of flows in the sorted list is approximately 1,000. Increasing the perturbation further results in near-total loss of locality, reducing the representation to behave similarly to conventional eviction strategies like LRU or LFU, with no spatial awareness. Based on this analysis, we limit the maximum perturbation to 30\%, balancing spatial diversity and information retention. This ensures that perturbed Bloom Filter states still encode meaningful locality cues to guide eviction decisions.

\section{Experimental Evaluation}

We comprehensively evaluate our proposed DQN-based eviction mechanism, demonstrating its efficacy in comparison to the traditional LFU and LRU flow eviction policies. We conduct extensive experiments across various parameters to thoroughly investigate and validate the benefits of our proposed solution.
All evaluations were conducted using a discrete-time simulator\footnote{\url{https://github.com/NWSL-UCF/DQN-Based-Flow-Management}} that models flow arrival, table occupancy, and eviction events in fine-grained time steps. This simulation framework allows us to explicitly define control intervals 
%such as RTI and ETI, 
along with other important parameters, and observe their impact on hit and miss behavior over time. 

\subsection{Parameter Exploration}

We extensively explore several key parameters to understand their impacts on performance. Table~\ref{tab:exp-params} summarizes the parameter ranges, while brief explanations of the most critical parameters are provided below for clarity:
\begin{itemize}
    \item \textbf{Reactive Time Interval (RTI):}  
    The latency (in seconds) required for the SDN controller to respond to a packet-in event by installing a new flow rule into the SFT.
    
    \item \textbf{Eviction Time Interval (ETI):}  
    The interval (expressed in multiples of RTI) after which our DQN-based eviction strategy periodically evaluates the occupancy of the SFT. For example, an ETI of 5 with an RTI of 0.01s translates into an eviction decision every 50 ms.
    
    \item \textbf{Idle Timeout:}  
    A standard timeout period (in seconds) after which flow entries with no observed traffic are automatically evicted from the SFT.
    
    \item \textbf{Packet Traces:}  
    Three distinct traces capturing heterogeneous traffic from various IoT devices ~\cite{unsw_iot_dataset}, offering diverse patterns of flow arrivals and activity levels.

    \item \textbf{DQN Learning Rate:}  
    Controls how quickly the DQN model updates its Q-values. Lower values lead to more stable but slower learning.

    \item \textbf{DQN Gamma (\(\gamma\)):}  
    The discount factor that balances immediate and future rewards in Q-value computation. A gamma close to 1 encourages long-term planning.

    \item \textbf{DQN Hidden Layers:}  
    The neural architecture of the Q-network, expressed as a sequence of hidden layer sizes.% used to approximate the Q-table.

\end{itemize}

\begin{table}[htbp]
\centering
\vspace{-3mm}
\caption{Explored Parameter Ranges for Evaluation}
\vspace{-2mm}
\begin{tabular}{|l|p{5.5cm}|}
\hline
\textbf{Parameter} & \textbf{Values Tested} \\ \hline
DQN Seeds & 101, 103, 107 \\ \hline
Table Size (SFT) & 16, 32 \\ \hline
Packet Traces & 3 traces: (600–1000 flows, 300k–500k packets) \\ \hline
RTI (s) & 0.01, 0.05 \\ \hline
Idle Timeout (s) & 30.0 \\ \hline
ETI (x RTI) & 5, 10, 100, 500 \\ \hline
DQN Learning Rate & 0.001, 0.01, 0.05, 0.1, 0.5 \\ \hline
DQN Gamma & 0.90, 0.99 \\ \hline
DQN Hidden Layers & \begin{tabular}[c]{@{}l@{}}32\_32\_32\_32\_32, 64\_64\_64\_64, \\ 128\_128\_128, 256\_256, 512\_512\end{tabular} \\ \hline
\end{tabular}
\label{tab:exp-params}
\vspace{-3mm}
\end{table}

% \subsection{Performance Analysis}
% \begin{figure*}[htbp]
%     \centering
%     \begin{subfigure}[b]{0.32\textwidth}
%         \centering
%         \includegraphics[width=\textwidth]{plots/absolute_hitrate_difference.pdf}
%         % \caption{\textbf{Hit rate difference (DQN+LRU vs LRU)}}
%         \caption{}
%         \label{fig:hit_rate_difference}
%     \end{subfigure}
%     \hfill
%     \begin{subfigure}[b]{0.32\textwidth}
%         \centering
%         \includegraphics[width=\textwidth]{plots/absolute_miss_rate_comparison.pdf}
%         % \caption{\textbf{Absolute miss rate comparison}}
%         \caption{}
%         \label{fig:absolute_miss_rate}
%     \end{subfigure}
%     \hfill
%     \begin{subfigure}[b]{0.32\textwidth}
%         \centering
%         \includegraphics[width=\textwidth]{plots/missrate_plot.pdf}
%         % \caption{\textbf{Normalized miss rate comparison}}
%         \caption{}
%         \label{fig:normalized_miss_rate}
%     \end{subfigure}
%     \caption{Performance comparison of caching algorithms: (a) Hit rate improvement of DQN+LRU over LRU on trace-3 in the last 5 hours, (b) Miss rates of all three methods in the last 10 hours, and (c) Normalized miss rates in the last 10 hours.}
%     \label{fig:performance_comparison}
% \end{figure*}

We conduct discrete-event simulations in Python to evaluate the effectiveness of our proposed eviction mechanism. Specifically, we have evaluated five flow eviction methods:
\begin{enumerate}
\item \textbf{LFU}: A baseline method employing the traditional LFU eviction policy, which removes flows that have generated the fewest number of packet matches.
\item \textbf{LRU}: Another baseline method employing the traditional LRU eviction policy, which removes flows that have not generated packets for the longest duration.
\item \textbf{DQN+LFU}: The enhanced approach combining LFU with our DQN-based eviction strategy (Sec.~\ref{sec:DQN+LRU}).
\item \textbf{DQN+LRU}: Another enhanced approach combining LRU with our DQN-based eviction (Sec. \ref{sec:DQN+LRU}).
 %, utilizing the optimal configuration identified earlier.
\item \textbf{Optimal}: The optimal strategy that evicts the flow whose next packet will arrive the farthest in the future.
%, taking advantage of full future knowledge available from the traffic trace. Since our testbed uses pre-recorded traces, we can determine exactly which rule each future packet will match. This strategy depends on three parameters in our discrete simulation setup: the size of the SFT, the Reactive Time Interval (RTI), and the Idle Timeout. While RTI and Idle Timeout are fixed across all optimal configurations, the SFT size is aligned with the corresponding baseline .
\end{enumerate}

% Through extensive evaluation across thousands of configurations, consuming over 540,000 GPU hours, we identify the optimal setup for our framework. The best configuration includes a 32-entry SFT, a 10~ms RTI, and an ETI of 100 (1 decision/sec). The DQN achieves optimal performance with a learning rate of 0.1, a discount factor ($\gamma = 0.99$), and a two-layer Q-network with 512 neurons per layer (512\_512).
Through extensive evaluation of thousands of configurations totaling over 540,000 GPU hours, we identify optimal setups for both LFU- and LRU-based methods. For \textbf{DQN+LFU}, the optimal configuration includes a 16-entry SFT, 10~ms RTI, 30~s Idle Timeout, ETI of 5~s, a DQN learning rate of 0.001, discount factor $\gamma = 0.99$, and a three-layer Q-network (128\_128\_128). For \textbf{DQN+LRU}, the best performance was achieved with a 32-entry SFT, 10~ms RTI, ETI of 100 (1 decision/sec), a learning rate of 0.1, $\gamma = 0.99$, and a two-layer Q-network (512\_512). For fair comparison, we configure the SFT size for \textbf{Optimal} to 16 entries when comparing to \textbf{DQN+LFU} and 32 entries when comparing to \textbf{DQN+LRU}.

\begin{figure}[htbp]
    \vspace{-3mm}
    \centering
    \includegraphics[width=0.40\textwidth]{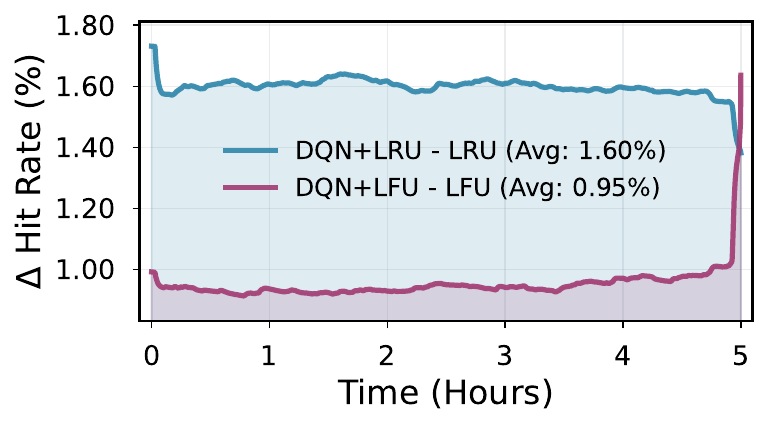}
    \vspace{-3mm}
    \caption{Hit rate improvement of DQN+LRU over LRU and DQN+LFU over LFU during the last 5 hours of trace-3}
    \label{fig:hit_rate_difference}
    \vspace{-3mm}
\end{figure}

\begin{figure}[htbp]
    \vspace{-3mm}
    \centering
    \includegraphics[width=\linewidth]{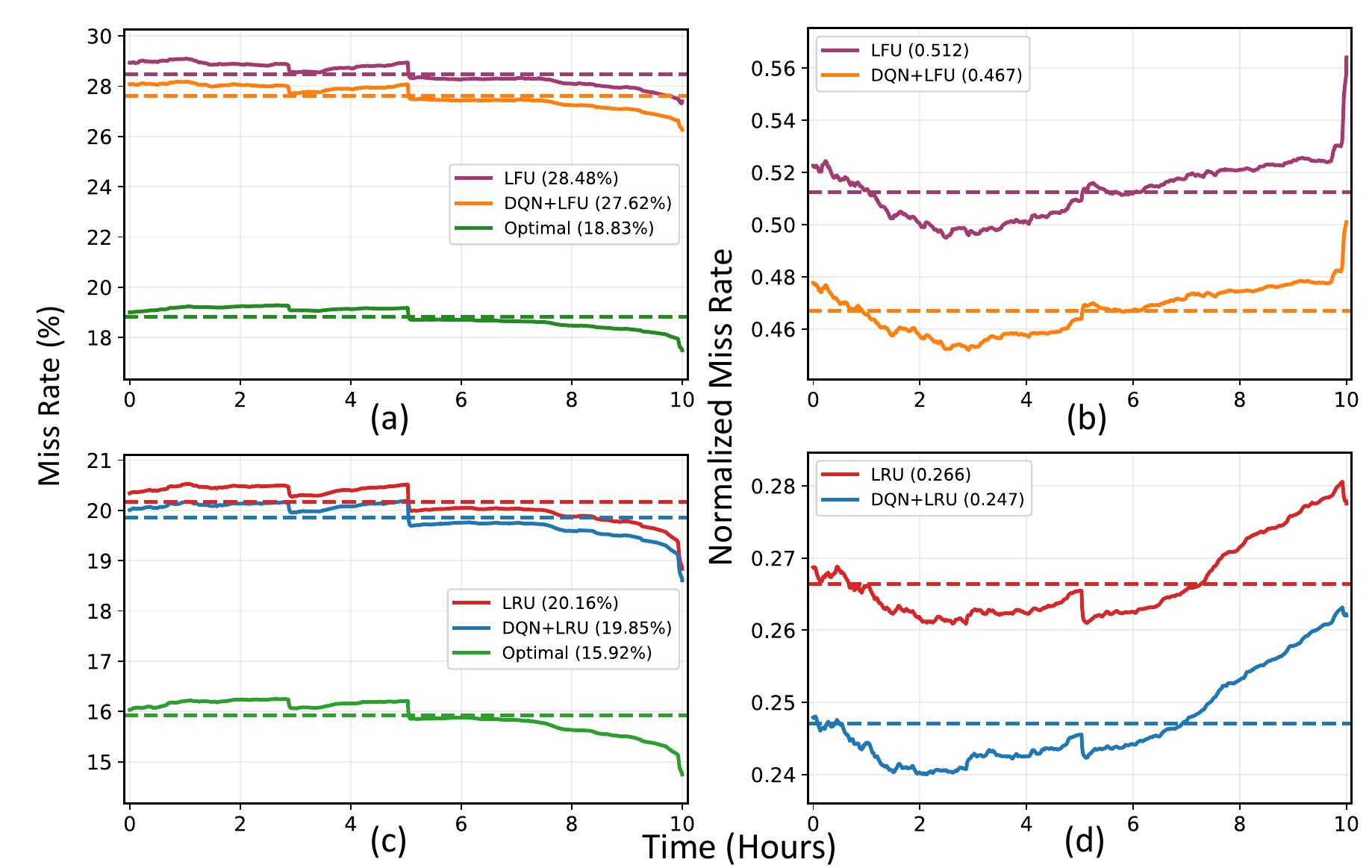}
    \vspace{-5mm}
    \caption{Miss Rate and Normalized Miss Rate}%between LFU and DQN+LFU, and between LRU and DQN+LRU}
    \label{fig:missrate}
    \vspace{-3mm}
\end{figure}

Our \textit{DQN-enhanced approach consistently maintains superior performance}, highlighting its ability to gradually learn and adjust to evolving traffic patterns. 
\textbf{Figure~\ref{fig:hit_rate_difference}} shows that DQN+LFU steadily attains higher hit rate than LFU, with an average of 0.95\% higher hit rate during the last 5 hours of trace-3. Similarly, DQN+LRU achieves 1.60\% higher hit rate than LRU over the same period. Such consistent improvement is particularly valuable for latency-sensitive applications, such as augmented reality (AR) and virtual reality (VR), where minimizing latency directly contributes to enhanced user experiences. Both \textbf{Figure~\ref{fig:missrate}a} and \textbf{Figure~\ref{fig:missrate}c} compare the miss rates of LFU, DQN+LFU, and Optimal eviction strategies, and LRU, DQN+LRU, and Optimal eviction strategy over the three traces during the last 10 hours. DQN+LFU and DQN+LRU again consistently improve the miss rate with respect to LFU and LRU, respectively, providing further evidence that our flowset encoding approach works well. Since LFU and LRU both cannot capture spatial locality, they yield worse performance against the Bloom Filter-encoded DQN-based method which can capture spatial locality among flows.

To better quantify the performance gap between traditional policies and DQN enhanced strategies, we define  \textit{normalized miss rate} as 
$
%\eta = \frac{M_{\text{strategy}} - M_{\text{optimal}}}{M_{\text{optimal}}}
(M_{\text{strategy}} - M_{\text{optimal}}) / M_{\text{optimal}}
$, where $M_{\text{strategy}}$ is to the miss rate of either LRU or DQN+LRU, and $M_{\text{optimal}}$ is the miss rate under the Optimal eviction method.
\textbf{Figure~\ref{fig:missrate}b} and \textbf{Figure~\ref{fig:missrate}d} show that integrating DQN with LFU consistently reduces the normalized miss rate, on average from 0.512 to 0.467—an approximate 8\% improvement over LFU, and from 0.266 to 0.247—an approximate 7\% improvement over LRU. This indicates that the DQN-enhanced policies \textit{reliably reduce the miss rate by bringing it closer to the Optimal}.

% \section{Summary and Future Work}
% \begin{itemize}
% \item Summary of key findings and their implications for SDN applications.
% \item Relevance of results in real-world network environments (data centers, IoT).
% \item Potential practical challenges or limitations.
% \item Specific future research directions (e.g., scalability, other RL techniques, deployment studies).
% \end{itemize}

\section{Summary and Future Work}

A key contribution of this work lies in the development of a scalable flow representation framework for RL in SDN. Traditional flow-aware designs often require maintaining large sets of flow identifiers, which become computationally and memory intensive for learning agents like DQN. To address this, we have proposed a Bloom Filter-based encoding that significantly compresses the flow space into a compact bit array, serving as an efficient state representation for the agent. This compression does not come at the cost of structural fidelity, and it preserves spatial locality among flows, allowing the agent to infer meaningful patterns. This spatial coherence was empirically validated through controlled perturbation experiments on the Bloom Filter, demonstrating its ability to retain locality-aware characteristics. 

To demonstrate the effectiveness of our Bloom Filter-based encoding in preserving spatial locality, we have designed a DQN-based flow caching framework in reactive SDN by using the full encoded flow space as the input state to the DQN agent. This design enables the agent to capture meaningful flow patterns while operating alongside the traditional policies, such as LRU and LFU, leading to a notable reduction in miss rate. By retaining spatial structure within a highly compressed representation, our approach facilitates intelligent, scalable decision-making without overwhelming the learning agent.

Future research can expand upon this work in several promising directions. First, investigating alternative reinforcement learning algorithms—such as Proximal Policy Optimization (PPO) or Asynchronous Advantage Actor-Critic (A3C)—may offer improved learning stability and robustness in decision-making. To further assess the real-world applicability of our framework, deploying it in production-scale SDN testbeds would provide valuable insights into its operational viability, including latency impact and fault tolerance.

\vspace{-2mm}
\bibliographystyle{IEEEtran}
\bibliography{references}

\begin{thebibliography}{10}
\providecommand{\url}[1]{#1}
\csname url@samestyle\endcsname
\providecommand{\newblock}{\relax}
\providecommand{\bibinfo}[2]{#2}
\providecommand{\BIBentrySTDinterwordspacing}{\spaceskip=0pt\relax}
\providecommand{\BIBentryALTinterwordstretchfactor}{4}
\providecommand{\BIBentryALTinterwordspacing}{\spaceskip=\fontdimen2\font plus
\BIBentryALTinterwordstretchfactor\fontdimen3\font minus \fontdimen4\font\relax}
\providecommand{\BIBforeignlanguage}[2]{{%
\expandafter\ifx\csname l@#1\endcsname\relax
\typeout{** WARNING: IEEEtran.bst: No hyphenation pattern has been}%
\typeout{** loaded for the language `#1'. Using the pattern for}%
\typeout{** the default language instead.}%
\else
\language=\csname l@#1\endcsname
\fi
#2}}
\providecommand{\BIBdecl}{\relax}
\BIBdecl

\bibitem{sdn_survey}
D.~Kreutz, F.~M.~V. Ramos, P.~E. Veríssimo, C.~E. Rothenberg, S.~Azodolmolky, and S.~Uhlig, ``Software-defined networking: A comprehensive survey,'' \emph{Proc. of IEEE}, vol. 103, no.~1, pp. 14--76, 2015.

\bibitem{openflow}
N.~Ruchansky and D.~Proserpio, ``A (not) nice way to verify the {OpenFlow} switch specification: Formal modelling of the {OpenFlow} switch using alloy,'' in \emph{Proc. of ACM SIGCOMM}, 2013, pp. 527--528.

\bibitem{tcam_limitations}
S.~Bhowmik, M.~A. Tariq, A.~Balogh, and K.~Rothermel, ``Addressing {TCAM} limitations of software-defined networks for content-based routing,'' in \emph{Proc. of ACM DEBS}, 2017, pp. 100--111.

\bibitem{yu2010scalable}
M.~Yu, J.~Rexford, M.~J. Freedman, and J.~Wang, ``Scalable flow-based networking with {DIFANE},'' \emph{ACM SIGCOMM CCR}, vol.~40, no.~4, pp. 351--362, 2010.

\bibitem{ddos_detection}
N.~Z. Bawany, J.~A. Shamsi, and K.~Salah, ``{DDoS} attack detection and mitigation using {SDN}: Methods, practices, and solutions,'' \emph{Arab. J. Sci. Eng.}, vol.~42, no.~2, pp. 425--441, 2017.

\bibitem{mice_elephant}
M.~Zaher, ``Traffic measurements, scheduling and characterization of {SDN}-based data center networks,'' Ph.D. dissertation, BUTE, Hungary, 2021.

\bibitem{rl_sdn}
Y.-R. Chen, A.~Rezapour, W.-G. Tzeng, and S.-C. Tsai, ``{RL}-routing: An {SDN} routing algorithm based on deep reinforcement learning,'' \emph{IEEE TNSE}, vol.~7, no.~4, pp. 3185--3199, 2020.

\bibitem{rl_sdn_2}
T.-Y. Mu, A.~Al-Fuqaha, K.~Shuaib, F.~M. Sallabi, and J.~Qadir, ``{SDN} flow entry management using reinforcement learning,'' \emph{ACM TAAS}, vol.~13, no.~2, pp. 1--23, 2018.

\bibitem{bloom1970space}
B.~H. Bloom, ``Space/time trade-offs in hash coding with allowable errors,'' \emph{Communications of the ACM}, vol.~13, no.~7, pp. 422--426, 1970.

\bibitem{bloom_routing}
A.~Marandi, T.~Braun, K.~Salamatian, and N.~Thomos, ``{BFR}: A bloom filter-based routing approach for information-centric networks,'' in \emph{IFIP Networking}, 2017, pp. 1--9.

\bibitem{bloom_ids}
B.~Groza and P.-S. Murvay, ``Efficient intrusion detection with bloom filtering in controller area networks,'' \emph{IEEE TIFS}, vol.~14, no.~4, pp. 1037--1051, 2018.

\bibitem{bloom_measurement}
A.~Kumar, J.~Xu, L.~Li, and J.~Wang, ``Space-code bloom filter for efficient traffic flow measurement,'' in \emph{P. of ACM IMC}, 2003, pp. 167--172.

\bibitem{lru_sdn}
E.-D. Kim, Y.~Choi, S.-I. Lee, and H.~J. Kim, ``Enhanced flow table management scheme with an {LRU}-based caching algorithm for {SDN},'' \emph{IEEE Access}, vol.~5, pp. 25,555--25,564, 2017.

\bibitem{lfu_sdn}
H.~Luo, W.~Li, Y.~Qian, and L.~Dou, ``Mitigating {SDN} flow table {OverFlow},'' in \emph{P. of IEEE COMPSAC}, vol.~1, 2018, pp. 821--822.

\bibitem{rl_dq_caching}
F.~Xu, F.~Yang, S.~Bao, and C.~Zhao, ``{DQN}-inspired joint computing and caching resource allocation approach for software defined information-centric {Internet-of-Things} network,'' \emph{IEEE Access}, vol.~7, pp. 61,987--61,996, 2019.

\bibitem{rl_dq_caching_2}
A.~Hariri, M.~Yuksel, and D.~Mohaisen, ``{RL}-based speculative installation of unseen flows in {SDNs} for low-latency applications,'' in \emph{Proceedings of IEEE ICMLCN}, 2024, pp. 250--256.

\bibitem{rl_actorcritic}
A.~Sharma, S.~Tokekar, and S.~Varma, ``Actor-critic architecture based probabilistic meta-reinforcement learning for load balancing of controllers in software-defined networks,'' \emph{Automated Software Engineering}, vol.~29, no.~2, p.~59, 2022.

\bibitem{rl_ppo_sdn}
J.~Wu and Z.~Zhu, ``Intelligent routing optimization for {SDN} based on {PPO} and {GNN},'' \emph{J. of Network \& Computer Apps.}, p. 104249, 2025.

\bibitem{flow_clustering_rl}
E.~H. Bouzidi, A.~Outtagarts, R.~Langar, and R.~Boutaba, ``Dynamic clustering of software defined network switches and controller placement using deep reinforcement learning,'' \emph{Computer networks}, vol. 207, pp. 108,852, 2022.

\bibitem{unsw_iot_dataset}
A.~Sivanathan, H.~H. Gharakheili, F.~Loi, A.~Radford, C.~Wijenayake, A.~Vishwanath, and V.~Sivaraman, ``Classifying {IoT} devices in smart environments using network traffic characteristics,'' \emph{IEEE TMC}, vol.~18, no.~8, pp. 1,745--1,759, 2018.

\end{thebibliography}
\end{document}